# Flat bands and distinct density wave orders in correlated Kagome superconductor CsCr$_3$Sb$_5$

Shuting Peng[1,#], Yulei Han[2,1,#], Yongkai Li[3,4,#], Jianchang Shen[1], Yu Miao[1], Yang Luo[1], Linwei Huai[1], Zhipeng Ou[1], Hongyu Li[1], Ziji Xiang[1], Zhengtai Liu[5], Dawei Shen[6], Makoto Hashimoto[7], Donghui Lu[7], Yugui Yao[3,4], Zhenhua Qiao[1,*], Zhiwei Wang[3,4,8,*] and Junfeng He[1,*]

[1]*Department of Physics and CAS Key Laboratory of Strongly-coupled Quantum Matter Physics, University of Science and Technology of China, Hefei, Anhui 230026, China*

[2]*Department of Physics, Fuzhou University, Fuzhou, Fujian 350108, China*

[3]*Centre for Quantum Physics, Key Laboratory of Advanced Optoelectronic Quantum Architecture and Measurement (MOE), School of Physics, Beijing Institute of Technology, Beijing 100081, China*

[4]*Beijing Key Lab of Nanophotonics and Ultrafine Optoelectronic Systems, Beijing Institute of Technology, Beijing 100081, China*

[5]*Shanghai Synchrotron Radiation Facility, Shanghai Advanced Research Institute, Chinese Academy of Sciences, Shanghai 201210, China*

[6]*National Synchrotron Radiation Laboratory and School of Nuclear Science and Technology, University of Science and Technology of China, Hefei 230026, China*

[7]*Stanford Synchrotron Radiation Lightsource, SLAC National Accelerator Laboratory, Menlo Park, California 94025, USA*

[8]*Material Science Center, Yangtze Delta Region Academy of Beijing Institute of Technology, Jiaxing, 314011, China*

[#]These authors contributed equally to this work.
*To whom correspondence should be addressed:
J.H.(jfhe@ustc.edu.cn), Z.W.(zhiweiwang@bit.edu.cn), Z.Q.(qiao@ustc.edu.cn)

**Kagome metal CsV$_3$Sb$_5$ has attracted much recent attention due to the coexistence of multiple exotic orders and the associated proposals to mimic unconventional high temperature superconductors. Nevertheless, magnetism and strong electronic correlations -- two essential ingredients for unconventional superconductivity, are absent in this V-based Kagome metal. CsCr$_3$Sb$_5$ is a newly discovered Cr-based parallel of CsV$_3$Sb$_5$, in which magnetism appears with charge density wave and superconductivity at different temperature and pressure regions. Enhanced electronic correlations are also suggested by theoretical proposals due to the calculated flat bands. Here, we report angle-resolved photoemission measurements and first-principles calculations on this new material system. Electron energy bands and the associated orbitals are resolved. Flat bands are observed near the Fermi level. Doping dependent measurements on Cs(Cr$_x$V$_{1-x}$)$_3$Sb$_5$ reveal a gradually enhanced band renormalization from CsV$_3$Sb$_5$ to CsCr$_3$Sb$_5$, accompanied by distinct spatial symmetry breaking states in the phase diagram.**

Multiple competing or intertwined orders may coexist in a material system when their associated energies are nearly degenerate. These orders can be selectively enhanced or suppressed by extrinsic parameters including doping, pressure or magnetic field, which in turn, provide a rich platform to investigate the underlying physics. In this context, Kagome metals AV$_3$Sb$_5$ (A=Cs, Rb, K) have attracted much attention due to the coexistence of many exotic orders, ranging from charge density wave (CDW) [1-11], superconductivity [1,2,12-14], nematic order [15,16], stripe order [17,18], pair density wave [18,19], and time-reversal symmetry breaking states [20-22]. The richness of the phase diagram has also stimulated proposals to link the AV$_3$Sb$_5$ to copper-oxide high temperature superconductors where similar orders have been observed [16,18-20,23,24]. However, magnetism and strong electronic correlations -- two essential ingredients for realizing the unconventional superconductivity [23,24], are clearly absent in AV$_3$Sb$_5$. Therefore, an important next step is to induce magnetism and electronic correlations to Kagome superconductors.

Recently, a huge amount of theoretical and experimental effort has been made to search for such candidates [25-29]. A potential breakthrough may take place due to the newly synthesized Kagome compound CsCr$_3$Sb$_5$ [30]. CsCr$_3$Sb$_5$ is a Cr-based parallel of CsV$_3$Sb$_5$ with a similar lattice structure [1,30,31], but it is energetically less stable than CsV$_3$Sb$_5$ [25]. Therefore, the successful growth of CsCr$_3$Sb$_5$ is highly nontrivial. More intriguingly, magnetism, charge density wave and superconductivity are reported at different temperature and pressure regions of the CsCr$_3$Sb$_5$ compound [30], and theoretical proposals suggest the existence of electronic correlation effects due to the calculated flat bands near the Fermi level [31,32]. The possible coexistence of magnetism and electronic correlation effects makes $CsCr_3Sb_5$ a unique platform for exploring unconventional electronic orders.

In this paper, we investigate the electronic structure of $CsCr_3Sb_5$ by angle-resolved photoemission spectroscopy (ARPES) measurements and first-principles calculations. Due to the technical challenges in sample growth, the size of the $CsCr_3Sb_5$ single crystals is still limited. By taking advantage of a small beam spot, we have successfully revealed the Fermi surface and orbital-resolved band structure of $CsCr_3Sb_5$ from the tiny single crystals. The overall electronic structure bears some resemblance to that of first-principles calculations, but renormalizations of the energy bands are clearly identified. Orbital-selective flat bands are observed near the Fermi level. Doping dependent measurements on $Cs(Cr_xV_{1-x})_3Sb_5$ reveal a monotonic enhancement of the band renormalization from $CsV_3Sb_5$ to $CsCr_3Sb_5$. Our results have also shed light on the nature of the symmetry breaking states in $CsCr_3Sb_5$ by unveiling the potential roles of structural instability and flat band induced electronic correlation.

Single crystals of $CsCr_3Sb_5$ were grown by a self-flux method using binary Cs-Sb eutectic mixture as a flux (see supplemental Fig. S1). The raw materials (Cs:Cr:Sb=7:3:14) were loaded into aluminum crucible and sealed in an evacuated quartz tube, which was heated slowly to 950 °C and stayed for 20 h. It was cooled down to 200 °C in 250 h and subsequently down to room temperature with the furnace switched off. Shiny crystals with apparent hexagonal edges were obtained after the flux was removed in distilled water. The ARPES measurements were performed at BL03U of Shanghai Synchrotron Radiation Facility and Stanford Synchrotron Radiation Lightsource (SSRL) beamline 5-2. The single crystals were cleaved and measured in-situ with a base pressure of better than $5 \times 10^{-11}$ torr. A small beam spot was used to probe the intrinsic electronic structure from the tiny fresh surface of the $CsCr_3Sb_5$ crystals. The Fermi level was obtained by measuring a polycrystalline gold piece in electrical contacts with the samples. First-principles calculations were performed by using the projected augmented-wave method [33] as implemented in the Vienna ab initio simulation package (VASP) [34,35]. The exchange-correlation interaction was addressed using the Perdew-Burke-Ernzerhof type of generalized gradient approximation [36]. The cut-off energy for plane wave basis and the energy convergence threshold were set to be 500eV and $10^{-6}$ eV, respectively. WANNIER90 [37] was used to construct a Wannier-based tight-binding model from the Cr-d and Sb-p orbitals, and the Fermi surface was obtained by utilizing the WannierTools package [38].

The crystal structure of $CsCr_3Sb_5$ is similar to that of $CsV_3Sb_5$ [Fig. 1(a)]. The three-dimensional (3D) Brillouin zone (BZ) and the projected in-plane BZ of $CsCr_3Sb_5$ are presented in Fig. 1(b). We use Γ (A), K(H), and M(L) to mark the specific momentum points in the 3D BZ, and use $\bar{\Gamma}$, $\bar{K}$ and $\bar{M}$ to denote the projected momentum points in the in-plane BZ, hereafter. The calculated and measured Fermi surfaces are shown in Fig. 1c and Fig. 1d, respectively. An overall agreement is achieved between the calculation and experiment. The Fermi surface consists of a big pocket centered at $\bar{\Gamma}$ and several smaller pockets centered at $\bar{M}$. Constant energy maps measured at selected binding energies are shown in Fig. 1e, where the momentum distribution of the photoelectron spectral weight exhibits characteristic patterns of the Kagome lattice.

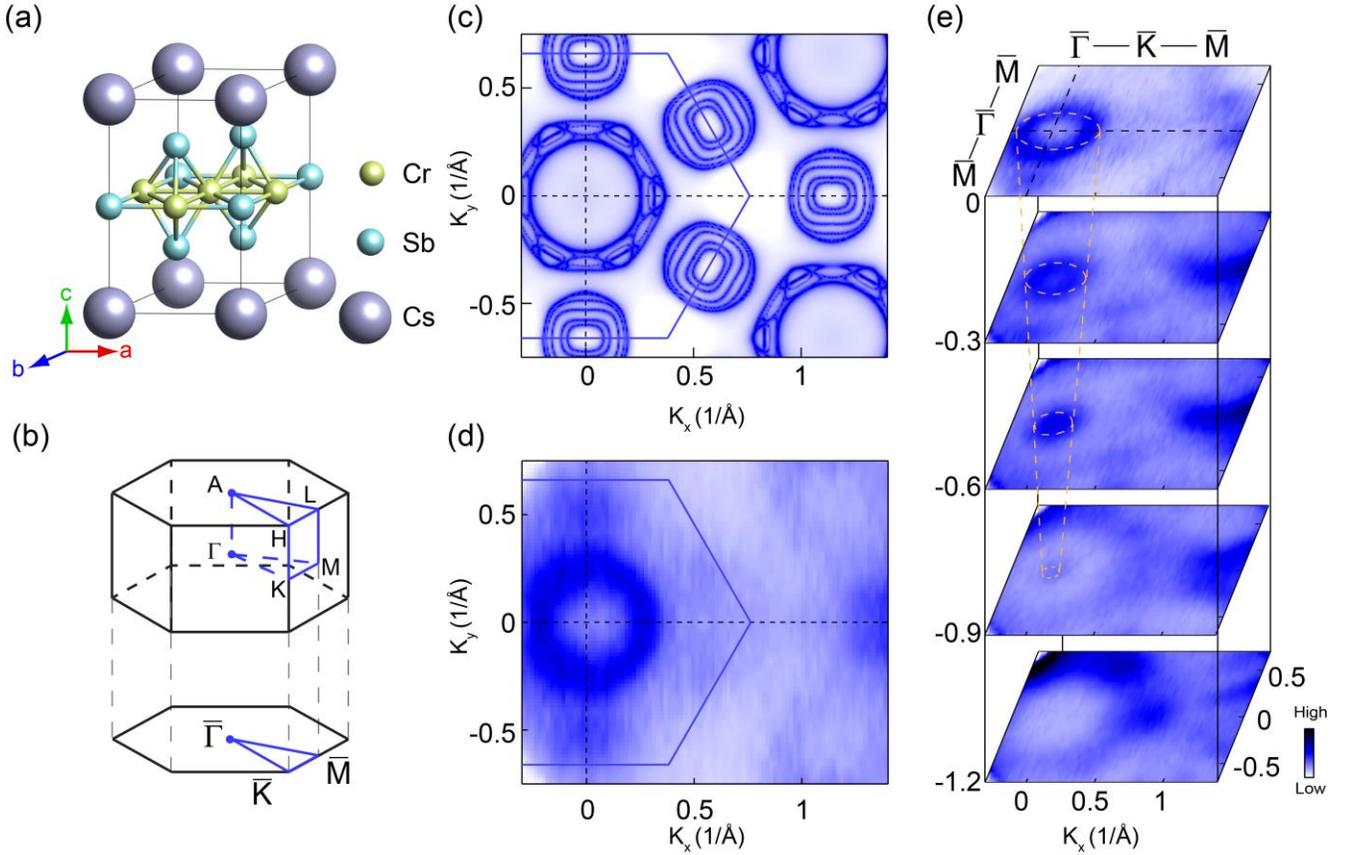

**FIG. 1. Crystal structure and electronic structure of $CsCr_3Sb_5$.** (a) Crystal structure of $CsCr_3Sb_5$. (b) Schematic of the 3D BZ and the projected in-plane BZ. (c) Calculated Fermi surface of $CsCr_3Sb_5$. (d) Fermi surface measured with 50 eV linear horizontally (LH) polarized light. (e) Constant energy maps at -0.3eV, -0.6eV, -0.9eV and -1.2eV with respect to $E_F$, respectively.

Fig. 2 shows the electronic structure of $CsCr_3Sb_5$ along high-symmetry directions. The calculated band structure retains typical features of a Kagome metal, including Dirac-like crossings, saddle points and flat bands (Fig. 2a,b and supplemental Fig. S2). In the meantime, special features of $CsCr_3Sb_5$ are also evident in the calculations. For example, the energy bands are formed by complex electron orbitals, and the calculated flat bands are not far from the Fermi level. Experimentally, the band structure of $CsCr_3Sb_5$ has been probed by photoemission measurements (Fig. 2 c-k). The overall dispersion shares some similarities with the calculation, but substantial band renormalization effects are discernible. Systematic measurements have also been carried out to examine the electron orbitals and flat bands predicted by the calculations. First, polarization-dependent photoemission measurements have been performed along both $\bar{\Gamma}$-$\bar{K}$-$\bar{M}$ and $\bar{\Gamma}$-$\bar{M}$ directions. Due to matrix element effects of photoemission, electron bands with different orbitals can be selectively probed by using different polarizations of the incident light. As shown in Fig. 2c,d, the bands contributed by Cr $d_{xz}$ and $d_{xy}$ orbitals along the $\overline{\Gamma}$-$\overline{K}$-$\overline{M}$ direction and the band with Cr $d_{yz}$ orbital along the $\overline{\Gamma}$-$\overline{M}$ direction are clearly observed by linear horizontally (LH) polarized light. The bands contributed by Cr $d_{yz}$ and $d_{x2-y2}$ orbitals along the $\overline{\Gamma}$-$\overline{K}$-$\overline{M}$ direction and the band with Cr $d_{xz}$ orbital along the $\overline{\Gamma}$-$\overline{M}$ direction are probed by linear vertically (LV) polarized light. These results are well aligned with the matrix element analysis based on the first-principles calculations (Fig. 2b). Nevertheless, the electron-like band centered at $\overline{\Gamma}$ shows discrepancies between the experiment and calculation. While the calculated Sb $P_z$ orbital is expected to be probed by LV polarized light, the measured band is enhanced by the LH polarized light, indicating the possible existence of orbital hybridization [29,39]. Second, fine scans have been performed near the Fermi level to search for the possible flat bands. As shown in Fig. 2e-h, nondispersive features are indeed observed over a large momentum region near the Fermi level, along both the $\overline{\Gamma}$-$\overline{K}$ and $\overline{\Gamma}$-$\overline{M}$ directions. These features can be quantified by nondispersive peaks on the raw energy distribution curves (EDCs) (Fig. 2f and h). We note that different portions of the nondispersive features are enhanced by different polarizations of the light, indicating a multi-orbital nature of the flat bands. These results are consistent with the first-principles calculations (Fig. 2b), but the energy positions of the flat bands are slightly lower than those in the calculation. As such, parts of the flat bands are observed below the Fermi level. Temperature dependent measurements do not reveal a significant change of the band structure, but the flat bands exhibit a small shift in energy at low temperature (Fig. 2 i-k).

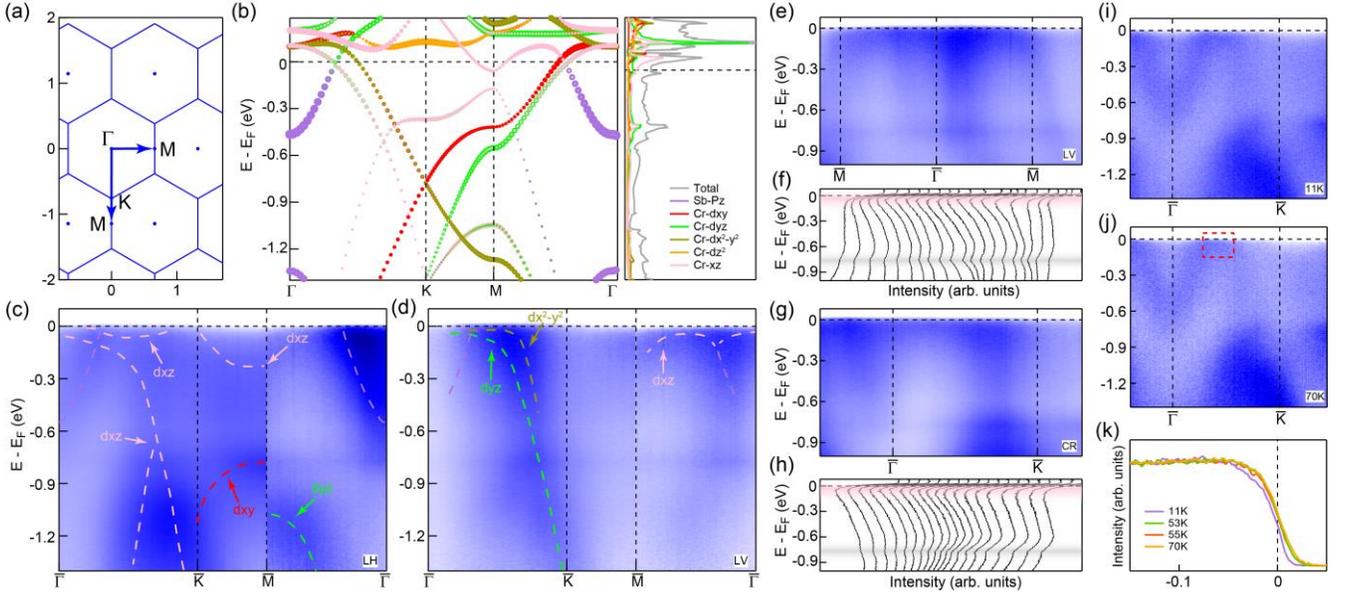

*Fig. 2 Polarization and temperature dependent measurements on $CsCr_3Sb_5$.* (a) The projected in-plane BZ and the momentum locations of the cuts. (b) Calculated orbital-resolved band structure and density of states (DOS) of $CsCr_3Sb_5$. Different orbitals are marked by different colors. The size of the markers represents the spectral weight of the orbitals. (c-d) Photoelectron intensity plots along $\bar{\Gamma}$-$\bar{K}$-$\bar{M}$-$\bar{\Gamma}$ of $CsCr_3Sb_5$ measured with 50 eV linear horizontally (LH) polarized (c) and linear vertically (LV) polarized (d) light. Dashed lines in (c), (d) are the eye-guides for bands with different orbitals. (e-h) Orbital-selective flat bands near the Fermi level measured with LV (e, f) and circularly (g, h) polarized light. (f, h) Raw energy distribution curves (EDCs) measured along $\bar{M}$-$\bar{\Gamma}$-$\bar{M}$ (f) and $\bar{\Gamma}$-$\bar{K}$ (h), respectively. The pink regions indicate the EDC peaks of the flat bands. The gray regions mark another flat feature at a deeper binding energy. (i-j) Photoelectron intensity plot measured at 11 K (i) and 70 K (j), respectively. (k) Momentum-integrated EDC in the red boxed area in (j) as a function of temperature.

After revealing the electronic structure of $CsCr_3Sb_5$, we now investigate the doping evolution from $CsV_3Sb_5$ to $CsCr_3Sb_5$. The experimental band structures (Fig. 3a-c) are compared with their corresponding calculation results (Fig. 3d-f). The discrepancies are quantified by the total energy differences at each momentum point (Fig. 3g-i). $CsV_3Sb_5$ shows clear and sharp bands (Fig. 3a), which are well captured by the calculations (Fig. 3d,g). Cr doping reduces the sharpness of the measured bands [$Cs(V_{0.55}Cr_{0.45})_3Sb_5$, Fig. 3b], and increases the discrepancies between the experimental and calculated band dispersion (Fig. 3h). The complete substitution of V by Cr further broadens the spectrum of the measured bands (Fig. 3c), and the extracted band dispersion shows substantial differences from that of the calculations (Fig. 3i). As such, these results reveal an increasing renormalization of the energy bands as a function of Cr doping.

The difference between $CsCr_3Sb_5$ and $CsV_3Sb_5$ is not only evident in their electronic structure but also in the presence of distinct symmetry breaking states (Fig. 4a). The electrical resistivity of $CsCr_3Sb_5$ exhibits a sharp peak at ~55K accompanied by a sudden curvature change (Fig. 4b, and supplemental Fig. S3). These features have been suggested as the charge density wave (CDW) and spin density wave (SDW) phase transitions [30]. CDW order has also been reported in $CsV_3Sb_5$, which manifests itself as a kink feature in the temperature dependent electrical resistivity curve (Fig. 4b). First, we discuss the SDW order that is unique to the Cr-based compound. We note that the measured band structure does not show a significant change upon entering the SDW state (Fig. 2i,j). The characteristic features of SDW (e.g. band splitting or band folding) are absent in the measurements. These results indicate that the magnetic order is either localized or fluctuating with a relatively weak magnitude. Then, we focus on CDW orders which appear in both $CsCr_3Sb_5$ and $CsV_3Sb_5$ but with different manifestations. The CDW in $CsV_3Sb_5$ exhibits an in-plane wavevector (Q) of 2X2, whereas the charge order in $CsCr_3Sb_5$ appears with a supermodulation of 1X4 (Fig. 4a). The Fermi surface map clearly shows the absence of the required nesting conditions in CsCr₃Sb₅ (Fig. 4a), which rules out Fermi surface nesting as the origin. Next, we consider the possible structural instability. Phonon spectra of CsCr₃Sb₅ have been calculated and shown in Fig. 4e. Similar to the case in CsV₃Sb₅, imaginary frequencies also appear in the phonon spectra of CsCr$_3$Sb$_5$, indicating the presence of structural instability. As established in CsV$_3$Sb$_5$, total energy is another window to examine the structural instability of the Kagome lattice [7,29,40]. In CsV$_3$Sb$_5$, both Star-of-David (SD) and Inverse-Star-of-David (ISD) structures are energetically more stable than the Kagome structure, which would naturally lead to a CDW phase transition with an in-plane wavevector of 2X2 (Fig. 4c). However, the Kagome lattice becomes more stable as a function of Cr doping (Fig. 4c), which echoes the weakening of the 2X2 CDW order with Cr doping (Fig. 4a and supplemental Fig. S3). The similar behavior is also seen in Ti and Ta doped CsV$_3$Sb$_5$ (see supplemental Fig. S4). It is clear that the Kagome structure has the lowest total energy among the Kagome, SD and ISD structures in CsCr$_3$Sb$_5$, which explains the absence of 2X2 CDW order. However, the consideration of 1X4 supermodulations in CsCr$_3$Sb$_5$ leads to a much lower total energy than that of the Kagome structure (Fig. 4d and supplemental Fig. S5). Therefore, structural instability represents a plausible driving mechanism of the 1X4 CDW in CsCr$_3$Sb$_5$. Finally, we discuss whether the electronic correlation in CsCr$_3$Sb$_5$ plays a role in the formation of the CDW order. In principle, the presence of flat bands near the Fermi level may significantly enhance the electronic correlation of the CsCr$_3$Sb$_5$ system and give rise to the CDW order. This scenario is in accord with our temperature-dependent measurements, where the flat bands exhibit a sudden energy shift across the phase transition temperature (Fig. 2k). The substantially enhanced electronic correlation in CsCr$_3$Sb$_5$ is also evidenced by the significant band renormalization effect as a function of Cr doping (Fig. 3 and Fig. 4f). We note that strong correlation effects might induce unconventional CDW orders, which can further compete or intertwine with SDW and superconductivity [23,24].

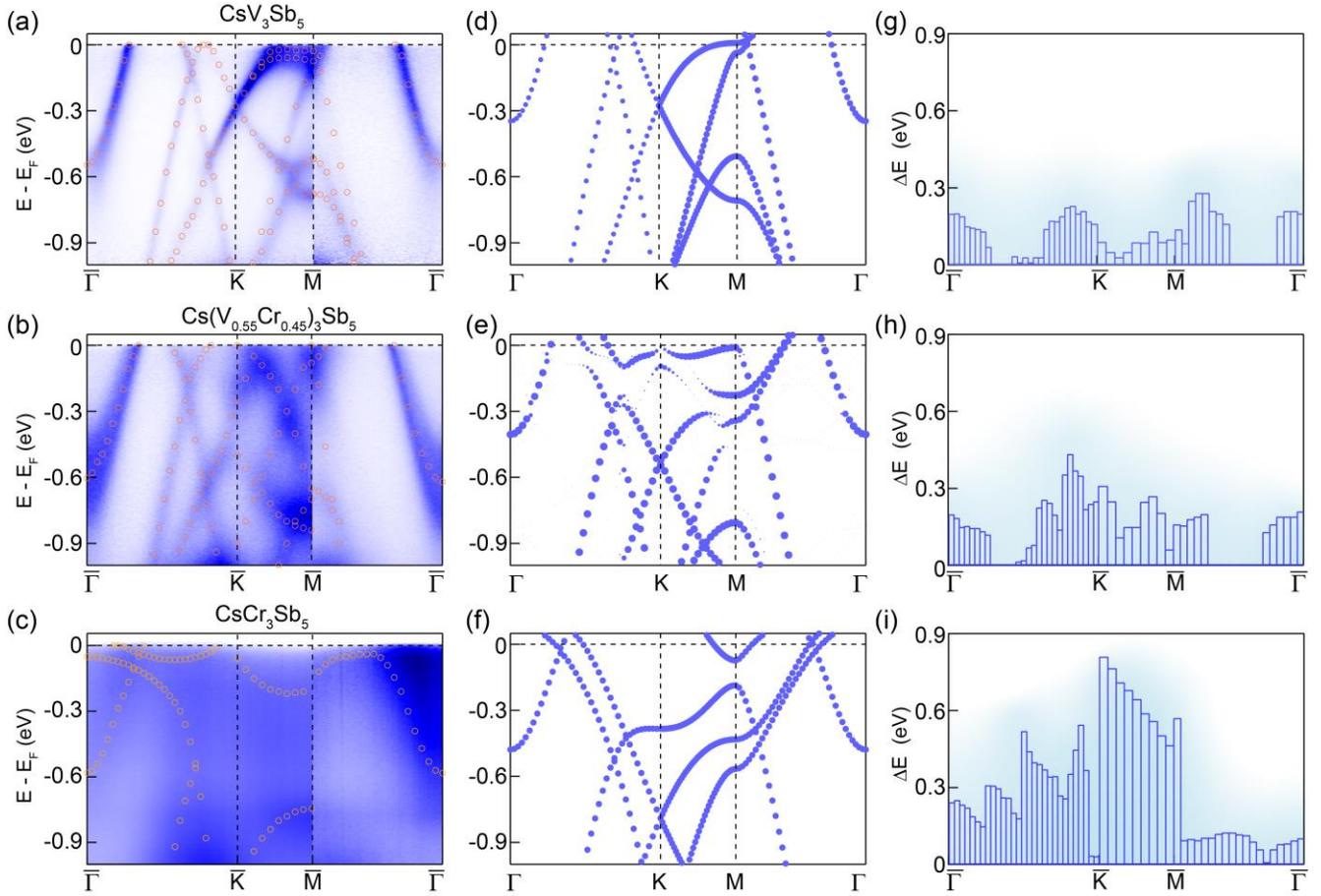

*Fig. 3 Evolution of the electronic structure with Cr doping.* (a-c) Photoelectron intensity plots along $\overline{\Gamma}$-$\overline{K}$-$\overline{M}$-$\overline{\Gamma}$ of $CsV_3Sb_5$ (a), $Cs(V_{0.55}Cr_{0.45})_3Sb_5$ (b) and $CsCr_3Sb_5$ (c). (d-f) Calculated band structure of $CsV_3Sb_5$ (d), $Cs(V_{0.5}Cr_{0.5})_3Sb_5$ (e) and $CsCr_3Sb_5$ (f). (g-i) The differences in energy as a function of momentum between the measured and calculated bands in $CsV_3Sb_5$ (g), $Cs(V_{0.55}Cr_{0.45})_3Sb_5$ (h) and $CsCr_3Sb_5$ (i).

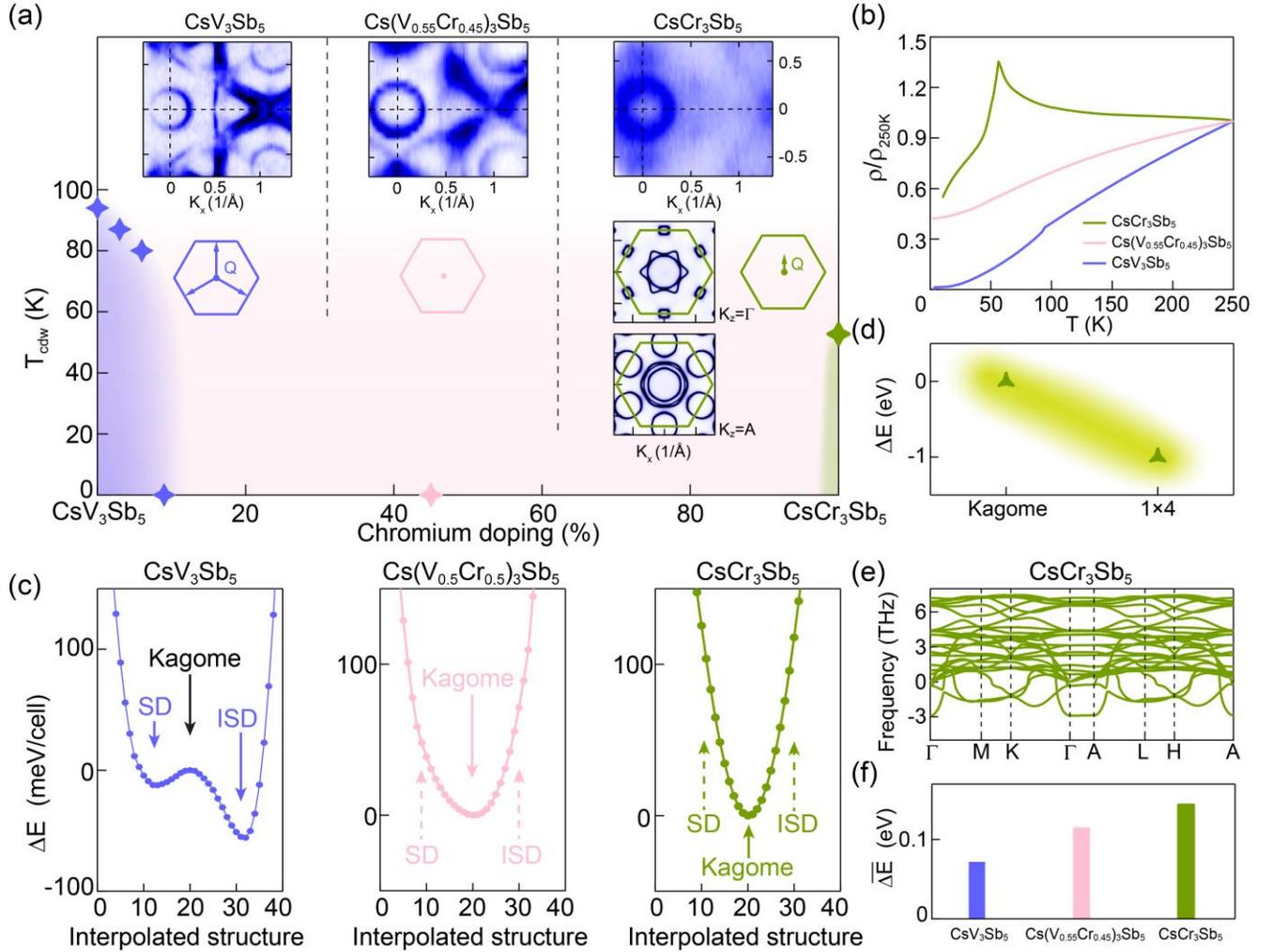

***Fig. 4 Origin of the CDW in CsCr₃Sb₅.*** *(a) CDW transition temperature ($T_{cdw}$) as a function of Cr doping. Fermi surface of CsV₃Sb₅, Cs(V₀.₅₅Cr₀.₄₅)₃Sb₅ and CsCr₃Sb₅ are shown in the insets. (b) Electrical resistivity of CsV₃Sb₅, Cs(V₀.₅₅Cr₀.₄₅)₃Sb₅ and CsCr₃Sb₅. (c) Total energy as a function of the interpolated structure in CsV₃Sb₅, Cs(V₀.₅Cr₀.₅)₃Sb₅ and CsCr₃Sb₅. ΔE stands for the relative total energy with respect to the Kagome structure per supercell. (d) Total energies of the Kagome structure and the structure with 1x4 supermodulations in CsCr₃Sb₅. (e) Calculated phonon spectra along the high-symmetry directions in CsCr₃Sb₅. (f) The averaged energy difference between the measured and calculated bands in CsV₃Sb₅, Cs(V₀.₅₅Cr₀.₄₅)₃Sb₅ and CsCr₃Sb₅.*

In summary, we have systematically investigated electronic structure of CsCr$_3$Sb$_5$ and identified the unique flat band features. Doping dependent measurements on Cs(V$_{1-x}$Cr$_x$)$_3$Sb$_5$ reveal a gradually enhanced band renormalization from CsV$_3$Sb$_5$ to CsCr$_3$Sb$_5$. Comparative studies suggest distinct symmetry breaking states in CsV$_3$Sb$_5$ and CsCr$_3$Sb$_5$, where the SDW in CsCr$_3$Sb$_5$ is presumably localized or fluctuating in nature, and the CDW in CsCr$_3$Sb$_5$ is associated with structural instability and electronic correlation effects. These results set CsCr$_3$Sb$_5$ apart from CsV$_3$Sb$_5$ and establish CsCr$_3$Sb$_5$ as a novel research platform to investigate the flat band induced electronic correlation as well as its manifestations in both spin and charge degrees of freedom.

We thank Xianhui Chen, Tao Wu, and Zhenyu Wang for useful discussions. The work at University of Science and Technology of China (USTC) was supported by the National Natural Science Foundation of China (No. 52273309, No. 12074358, No. 52261135638), the Fundamental Research Funds for the Central Universities (No. WK3510000015), the Innovation Program for Quantum Science and Technology (No. 2021ZD0302802), National Key R&D Program of China (Grants No. 2023YFA1406304) and National Science Foundation of China (Grant No. U2032208). The work at Beijing Institute of Technology was supported by the National Key R&D Program of China (Grants No. 2020YFA0308800, No. 2022YFA1403400), the Natural Science Foundation of China (Grant No. 92065109), the Beijing Natural Science Foundation (Grants No. Z210006, No. Z190006). Use of the Stanford Synchrotron Radiation Lightsource, SLAC National Accelerator Laboratory, is supported by the U.S. Department of Energy, Office of Science, Office of Basic Energy Sciences under Contract No. DE-AC02-76SF00515. M. H. and D. L. acknowledge the support of the U.S. Department of Energy, Office of Science, Office of Basic Energy Sciences, Division of Material Sciences and Engineering. J. H. thanks the Analysis & Testing Center at USTC for the support. Z.W. thanks the Analysis & Testing Center at BIT for assistance in facility support.